# Variable-wavelength quick scanning nano-focused X-ray microscopy for *in situ* strain and tilt mapping

*Marie-Ingrid Richard[†,‡,*], Thomas W. Cornelius[†], Florian Lauraux[†], Jean-Baptiste Molin[§], Christoph Kirchlechner[§], Steven J. Leake[‡], Jérôme Carnis[†,‡], Tobias U. Schülli[‡], Ludovic Thilly[⊥], Olivier Thomas[†]*

Dr. Marie-Ingrid Richard
Aix Marseille Université, CNRS, Université de Toulon, IM2NP UMR 7334, 13397, Marseille, France
ID01/ESRF, The European Synchrotron, 71 Avenue des Martyrs, Grenoble 38043 Cedex, France
E-mail: mrichard@esrf.fr

Dr. Thomas W. Cornelius
Aix Marseille Université, CNRS, Université de Toulon, IM2NP UMR 7334, 13397, Marseille, France

Florian Lauraux
Aix Marseille Université, CNRS, Université de Toulon, IM2NP UMR 7334, 13397, Marseille, France

Jean-Baptiste Molin, Dr. Christoph Kirchlechner
Max-Planck-Institut für Eisenforschung GmbH, Max-Planck-Strasse 1, 40237 Düsseldorf, Germany

Dr. Steven J. Leake
ID01/ESRF, The European Synchrotron, 71 Avenue des Martyrs, Grenoble 38043 Cedex, France

Dr. Jérôme Carnis
Aix Marseille Université, CNRS, Université de Toulon, IM2NP UMR 7334, 13397, Marseille, France
ID01/ESRF, The European Synchrotron, 71 Avenue des Martyrs, Grenoble 38043 Cedex, France

Dr. Tobias U. Schülli
ID01/ESRF, The European Synchrotron, 71 Avenue des Martyrs, Grenoble 38043 Cedex, France

Prof. Ludovic Thilly
Institut Pprime, UPR 3346, CNRS, University of Poitiers, ISAE-ENSMA, SP2MI, Boulevard Marie et Pierre Curie, BP 30179, 86962 Futuroscope Chasseneuil Cedex, France

Prof. Olivier Thomas
Aix Marseille Université, CNRS, Université de Toulon, IM2NP UMR 7334, 13397, Marseille, France

Compression of micro-pillars is followed *in situ* by a quick nano-focused X-ray scanning microscopy technique combined with three-dimensional reciprocal space mapping. Compared to other attempts using

X-ray nanobeams, it avoids any motion or vibration that would lead to a destruction of the sample. The technique consists of scanning both the energy of the incident nano-focused X-ray beam and the in-plane translations of the focusing optics along the X-ray beam. Here, we demonstrate the approach by imaging the strain and lattice orientation of Si micro-pillars and their pedestals during *in situ* compression. Varying the energy of the incident beam instead of rocking the sample and mapping the focusing optics instead of moving the sample supplies a vibration-free measurement of the reciprocal space maps without removal of the mechanical load. The maps of strain and lattice orientation are in good agreement with the ones recorded by ordinary rocking-curve scans. Variable-wavelength quick scanning X-ray microscopy opens the route for *in situ* strain and tilt mapping towards more diverse and complex materials environments, especially where sample manipulation is difficult.

**Introduction**

Strain and lattice distortions often dictate both performance and properties of materials.[1] A wide range of techniques have been developed to map strain and they can be divided into imaging and diffraction techniques. For example, in transmission electron microscopy (TEM), atomic resolution can be achieved by imaging atomic columns, which directly determines the local elastic strain from the observed lattice distortions.[2,3] However, it is more common for diffraction techniques to be used to measure the local elastic strain, where the Bragg law directly links the position of diffraction peaks to the lattice constants and is exploited with many different probes; X-ray,[4] neutron,[5] and electron diffraction.[2] Complex strain distributions require that the local strain is measured with high spatial resolution across the entire field-of-view. However, most of the time, the local strain is difficult to measure under real working conditions, in particular, at the nanoscale. Nano-focused Bragg (coherent) X-ray diffraction methods (see for example, Refs.[6–11]) are often used at synchrotron sources to address this challenge. These techniques allow one to image the strain fields at the nanoscale non-destructively and under complex sample environments thanks to the high penetration of X-rays in matter. Focal spot sizes of a few hundred nanometers are routinely obtained nowadays, rendering it possible to study individual nanostructures.

In order to fully record the induced structural changes during operation (such as at high temperatures, pressures, electric or magnetic fields *etc*), it is necessary to measure *in situ* the complete three-dimensional (3D) intensity distribution of a Bragg reflection coming from the nanostructure under investigation. 3D reciprocal space maps (3D-RSMs) are generally recorded by rocking the sample on the order of one degree or more. The typical sphere of confusion of state-of-the-art diffractometers is of the order of 10 micrometers over a complete rotation. This may lead to parasitic displacements caused by *e.g.* the eccentricity error of the rotation axis that can be much larger than both the beamsize and the nanostructures, thus complicating the measurement of 3D-RSMs. Moreover, the presence of complex sample environments may limit the sample movement or access in angular space and make it necessary to reduce or avoid any vibrations induced by sample rotation. As an example, during *in situ* compression tests using a micro-compression device or an atomic force microscope[12,13] to study the mechanical properties of single micro- and nanostructures, any movement of diffractometer motors must be avoided to prevent damaging the tip and/or nanostructure during loading. Therefore, a method of recording 3D-RSMs at different sample positions and without moving the sample is highly desirable. Alternatively, the X-ray energy can be varied in a pre-determined range, which allows for mapping reciprocal space in three dimensions when using a two-dimensional (2D) detector. Recent studies have successfully measured 3D-RSMs from crystals in this manner for a given sample position.[14–17]

Here, we use this method not for one sample position only but during 2D direct space mapping by scanning the focusing optics, hence translating the focused beam on the sample surface, while keeping the sample itself fixed. This approach of scanning the focusing optics is similar to the Kirkpatrick-Baez (KB) scanning method demonstrated for *in situ* Laue microdiffraction studies.[18] This variable-wavelength quick scanning Bragg X-ray microscopy technique thus allows to record scattering images of extended objects without moving any diffractometer motor. We will demonstrate that chromatic optics, namely Fresnel zone plates, can be successfully used for this purpose. They eliminate the need to rotate or translate the sample. 3D-RSMs are recorded by scanning the wavelength of the incident X-ray beam (as opposed to scanning the sample angle) at different direct space positions (*y* and *z*) of the sample (*x* being along the beam direction). The (*y*, *z*) maps are obtained by scanning the focusing optics, while keeping the sample fixed. This five-dimensional data set (3 reciprocal space coordinates and two dimensions in direct space) is processed and analysed using the X-ray strain orientation calculation software (XSOCS)[8] that we modified to take into account the variation of the X-ray energy. The resulting reciprocal space coordinates of the diffraction peak as well as the strain (resolution of $10^{-5}$)[19] and lattice tilt are compared with the same map obtained by the classical approach of rocking the Bragg angle and scanning the sample. This capability enables new strain imaging studies in environments where samples cannot move and the details of nanoscale strain distribution and evolution remain difficult to follow.

**Experimental method**

The experiment was performed at the upgraded ID01 beamline of the European Synchrotron Radiation Facility (ESRF). Monochromatic X-rays of an energy of 8 keV (wavelength λ of 1.54975 Å) are selected by a double crystal Si(111) monochromator. 3D-RSMs were recorded both by rocking scans and by scanning the energy of the incident X-ray beam. While in the former case, the sample was rotated by ± 0.1° (corresponding to 0.116 nm$^{-1}$ in reciprocal space), the X-ray energy was varied in the latter case by ΔE = ± 24 eV in steps of 6 eV corresponding to a variation of ΔQ = ± 0.098 nm$^{-1}$. To remain on the maximum of the undulator emission peak and, thus maintaining the incident intensity constant, the undulator gap was adjusted at every energy step.

The X-rays were focused by a Fresnel zone plate (FZP) having a diameter of 300 µm and an outermost zone width of 60 nm. Unfocused radiation transmitted by the FZP was eliminated using a 60 µm wide (80 µm thick) Au central stop placed in front of the FZP, while higher diffraction orders were eliminated by a 50-µm diameter Pt/Ir order sorting aperture (OSA) positioned 20 mm upstream of the sample (see **Figure 1**). The FZP is a chromatic focusing optics, *i.e.* its focal length $f$ varies as a function of the X-ray energy: $f(E) \propto E$. At 8 keV, the focal length of the FZP is 116 mm. A set of slits located in front of the FZP were closed down to an aperture of 200 µm (vertically) and 60 µm (horizontally) corresponding to the coherent part of the X-ray beam at the ID01 beamline. The focal profile of the beam was characterised using a 2D ptychography approach on a test pattern featuring a 30-µm diameter gold Siemens star placed close to the focal position of the FZP. The gold structure and the complex-valued wavefront were retrieved simultaneously using the ptychography reconstruction code of the PyNX package.[20] **Figure 2**(a) – (c) display cuts of the retrieved experimental complex illumination. This leads to a Gaussian beam waist with focal spot sizes of 80 nm (vertical) and 280 nm (horizontal) with respective Rayleigh ranges of 0.15 mm (vertical) and 1.59 mm (horizontal). The experimental and simulated 1D illumination profiles are displayed in Figure 2(d). Despite the asymmetry of the experimental profile, which is probably caused by the optical aberrations of the FZP, both curves lead to the same focal depth of ~ 0.33 mm (full width at half maximum of the peak size). Note that the focal depth is increased when reducing the illuminated area of the FZP.[21] Thus, the energy of the incident X-ray beam can be changed within a wider range without deteriorating the focal spot on the structure under investigation. Considering the depth of focus and linear chromaticity of the FZP, one can thus tolerate an energy variation of about ±0.2 % (+/- 16 eV) without significant beamsize variation in the vertical and of ±1.3 % (+/- 104 eV) for constant horizontal beamsize.

Free-standing Si [110]-oriented micro-pillars were fabricated on top of a Si wedge by focused ion beam (FIB) milling (inset of Figure 1). This geometry favours X-ray transmission thus facilitating the alignment of the nanostructure with respect to the nano-focused X-ray beam for *in situ* nano-focused quick X-ray microscopy. During the milling process, the ion beam current and acceleration voltage were successively reduced to limit eventual creation of lattice defects and/or surface amorphisation. During the last fabrication step when defining the pillar, ion beam currents of 24 pA as well as an ion energy of 30 keV were employed.

The sample was mounted on the load-cell of a custom-built micro-compression device,[22,23] which was installed on the hexapod (resolution of 100 nm) of the high precision diffractometer. A photograph of the experimental setup is shown in Figure 1. A flat diamond tip ($d$ = 10 µm) was used for *in situ* compression tests. An optical microscope was used to observe the sample and the tip from above, therefore allowing for accurate sample positioning at the center of rotation of the goniometer. The FZP was mounted on a piezoelectric stage (from Physik Instrumente) with a stroke of 100 µm along $x$, $y$ and $z$ and an encoder resolution of 2 nm. This mounting facilitates laterally scanning the focusing optics (along the $y$ and $z$-directions) and, hence the beam focus across an immobile sample, thus allowing for quick mapping without damaging the tip and/or the nanostructure under investigation.

The [110]-oriented Si pillars were mounted so that their axis was along the $y$-direction (see **Figure 3**). The (220) vertical planes were probed in a horizontal scattering geometry to get access to the axial strain of the Si pillars during their mechanical loading and unloading. The X-ray intensity around the **220** Si Bragg reflection of Bragg angle, $\theta_B$ = 23.8°, was recorded using a 2D MAXIPIX[24] fast read-out photon counting (frame rates of up to 100 Hz full frame) detector of 516×516 pixels and 55 µm pixel size, mounted at a distance of 1.4 m from the sample. The acquisition is performed continuously or "on the fly", *i.e.* with a very low detector deadtime (290 µs) and zero settling time for the motors.

**Results and discussion**

Figures 3(a-b) display two-dimensional direct space maps of the logarithmic integrated intensity measured at the **220** Si Bragg reflection for different surface areas (from large to small scale) centered around a Si micro-pillar (square section of 0.81 x 0.81 µm² and height of 2.26 µm) at the top of a large cubic pedestal (10 × 10 × 10 µm³) on the top of a Si wedge. Note that the maps were recorded by scanning the focusing optics in the *y-z* plane, whilst the sample remained stationary. The pillar appears smaller in height as it has been rotated in the plane (along the out-of-plane axis, φ) to be in Bragg condition. The scattered signal from the Si pedestal (see Figure 3(b)) shows heterogeneities indicating variations of strain and/or tilt of lattice planes.

The strain and lattice orientation of a pristine pillar (section of 0.92 µm x 0.92 µm and height of 2.82 µm - similar to the one shown in Figure 3) and of its pedestal were measured using the conventional quick scanning technique by finely scanning the Bragg angle (φ) at the energy of 8 keV (see **Figures 4 and 5**). A sample area of 5 µm × 5 µm was mapped taking diffraction with the 2D detector every 100 nm for different in-plane incident angles (27.54° ≤ φ < 27.74°) with steps of 0.01°. The angular resolution depends on the diffractometer (typically 0.0005° at the ID01 beamline of ESRF) and the convergence of the incident beam. The latter was 0.01° in the diffraction plane in this experiment. This resulted in the collection of 20 (*y*, *z*) maps (for 20 incidence angles). The lateral spatial resolution (here, along *y* and *z*) is given by the beam size. The resolution in the orthogonal direction (here, *x*) depends on the beam penetration depth (*i.e.*, energy and incident/exit angles of the X-ray beam and the studied material). The obtained detector frames at each spatial (*y*, *z*) position were then reconstructed into three-dimensional (3D) reciprocal space maps by converting the detector pixels and the angular coordinates into the reciprocal space coordinates, $Q_x$, $Q_y$ and $Q_z$. Figure 4 displays the reciprocal space coordinates of the scattering vector, $Q_x$ (a), $Q_y$ (b) and $Q_z$ (c). Figure 5 displays the maximum measured intensity (a) and the strain along the [220] direction, $\varepsilon_{220}$ (b). $\varepsilon_{220}$ is related to the variation of the $d_{220}$ spacing of the atomic planes, and was retrieved for each position in direct space: $\varepsilon_{220} = (d_{220,meas} - d_{220,ref})/d_{220,ref}$, where $d_{220,ref}$ is the reference *d* spacing of bulk Si. Interestingly, the top of the Si pillar exhibits compressive strain ($\varepsilon_{220}$ < 0) of the vertical (220) planes. On the contrary, one topside of the Si pedestal shows tensile strain ($\varepsilon_{220}$ > 0). Assuming that the mosaic spread was homogeneously distributed in all directions, the lattice tilt of the (220) Si planes was calculated from the peak position as $\eta = \arccos(Q_y/Q_{220})$ (see Figure 5(f)) revealing a tilt of up to 0.07° at the base of the Si pillar while in the interior of the pedestal no tilt was observed. Both the strain and the tilt of the (220) Si lattice planes in the Si pillar probably originate from the FIB-milling process.

The same sample area of 5 x 5 µm² was then mapped with the same spatial resolution as before but using the variable-wavelength quick scanning technique. The energy of the incident X-ray beam was varied in steps of 6 eV and the undulator gap was adjusted accordingly while keeping the rocking angle φ fixed for each fast (*y*, *z*) map. This resulted in the collection of 9 (*y*, *z*) maps (for 9 energies from 7.976 keV to 8.024 keV). **Figures 4 and 5** display the results obtained using the variable-wavelength quick scanning technique. They are in excellent qualitative and quantitative (the same scale bars are used for the strain and tilt in Figure 5) agreements with the ones obtained using the conventional quick scanning technique. The difference between the strain and tilt maps obtained by conventional and variable-wavelength quick scanning techniques can be estimated using the mean square error: $\varepsilon_r = \sum_{i,j}(I_{i,j}^1 - I_{i,j}^2)^2 /N$, where $I_{i,j}^{1,2}$ are the different pixels of image 1 or 2, which corresponds to the strain or tilt map obtained using the two techniques and *N* is the total number of pixels, respectively. The mean square errors for the strain and tilt maps between the two techniques are $\varepsilon_r = 1.4\times10^{-7}$ and $\varepsilon_r = 3.8\times10^{-5}°$. These values can be compared with the averaged values of the strain and tilt: -4.1×10⁻⁵ and 2.94×10⁻²°, respectively. This validates the variable-wavelength quick scanning X-ray microscopy approach.

The strain and tilt-sensitive imaging capability of this new quick scanning technique, which does not require any sample motion, was then employed during *in situ* mechanical loading, where the traditional rocking approach must not be used due to vibrations induced by moving diffractometer motors. **Figures**

**6 (a, b)** display the strain and lattice orientation fields of another Si pillar (shown in Figure 3 and different from Figure 5). Interestingly, its strain and lattice tilt are different from the ones shown for the pillar displayed in Figure 5. Different residual strain and lattice tilt are obtained from the FIB milling process from one pillar to the other. Figures 6 (c, d) display the strain and lattice orientation fields when the indenter is in contact with the Si pillar. Note that the same colorscale is used at the initial state (see Figures 6 (a, b)) and during contact (see Figures 6 (c, d)) for the strain and tilt, respectively. Small changes of the strain and tilt inside the Si pillar are observed when the tip is in contact with it. Figures 6(e-f) display the differences of strain and tilt between the contact and the initial state. It appears that the (200) planes at the top of the pillar are slightly compressed ($\Delta\varepsilon_{220}<0$) and tilted when the tip is in contact. When the tip is in contact, only the variable-wavelength quick scanning X-ray microscopy approach can be used whereas the traditional rocking approach risks to breaking the pillar or the tip. **Figure 7** displays the strain and lattice orientation fields of the Si pillar and the pedestal (shown in Figures 3 and 6) before (a-b) and during loading (c-d), where a force of 0.9 mN was applied corresponding to a stress of 1.37 GPa. A compressive strain of -0.81% was expected considering the Si Young's modulus of 169 GPa along [110]. Because of elastic bending of the Si micro-pillar, this maximum strain value was outside of the scanned reciprocal space range (covering $\Delta Q = \pm 0.123$ nm$^{-1}$); the Si pillar is thus not observed during loading. The *in-situ* variable-wavelength quick scanning map however evidences an inhomogeneous strain field in the Si pedestal with large compressive strain of up to -0.1 % just below the Si micro-pillar. This inhomogeneous compressive strain field results from the fact that the pillar is pushed into the Si pedestal, thus indenting it. Besides the strain, the maps show strong tilting of the Si (220) lattice planes in the pedestal ranging from -0.04 to +0.06° (compared to 0.02° in the pristine state). While negative lattice tilt values are revealed just below the Si nanopillar, this is compensated by a rotation of the lattice planes into the opposite direction about 2 µm below the top surface of the pedestal. Moreover, the rotation of the (220) planes spreads few micrometers away from the indented Si pillar.

Hence, as demonstrated in the present work, this new approach based on variable energy and on scanning the focal spot provides access both to the rotation of the crystalline lattice and to the strain field inside micro- and nanostructures under mechanical load offering new opportunities to study the mechanical behavior at small scales *in situ*. It calls for the development of precision mechanics for optics positioning in order to be able to move with high reproducibility the optics in three dimensions in order to adapt the focal distance to the X-ray energy and hence to overcome the limiting parameter of a narrow depth of field. By adapting the focal distance, monochromator and undulator gap, it is possible to enlarge the scanning energy range. With the improvement of compact and lightweight nano-optics with short focal lengths, this is likely to play an even more important role in the future. The very light weight of such optics however will likely make a precise positioning and scanning of these devices a standard option on nano-focused X-ray beamlines. Besides chromatic also achromatic focusing optics such as Kirkpatrick-Baez mirrors may be used. The installation of a hexapod underneath each mirror allows for translating them laterally and vertically and thus enables a two-dimensional displacement of the focused X-ray beam and, thus, the two-dimensional cartography of the sample. This kind of KB-scanning method has recently been demonstrated for Laue microdiffraction [18].

In comparison, among all the techniques to measure strain (for instance, spatially resolved Raman spectroscopy,[25,26] digital image correlation (DIC),[27] …), high-resolution electron backscatter diffraction (HR-EBSD) offers a strain and angular resolution of $10^{-4}$ and 0.006°, respectively, for *ex situ* experiment.[28,29] But the sample must be compatible with the general requirements of electron microscopy. Interestingly, *in situ* TEM deformation and characterization have been successfully carried out.[30,31] *In situ* TEM is characterized by an excellent spatial resolution of ~1 nm compared with ~100 nm (beam size) for X-ray microscopy. However, the strain resolution (~$10^{-3}$[32]) as well as the field of view (~150 x 150 nm$^2$ for high-resolution (HR)-TEM[33]) are several order of magnitude smaller than for X-ray microscopy (~10$^{-5}$[19] and ~200 x 200 µm$^2$). Note that the strain resolution is ~2.5x10$^{-4}$ in this work as we acquired (*y*, *z*) maps using a 2D detector each 6 eV step. The strain resolution can be improved by reducing the energy step size but at the expense of a longer measurement time. Furthermore, the TEM specimen preparation technique is often destructive so that the studied sample will not be in the same strain-state as the original "bulk" material. For HRTEM, lenses are not perfect, even in the age of aberration correction. There are therefore slight shifts between the image of the atomic lattice and its actual position.[34] Indeed, the position of neighboring atomic columns can even influence the apparent position of the atomic column of interest.

**Conclusion**

In summary, we introduced an *in situ* combination of mechanical loading and quick scanning nano-focused X-ray microscopy based on variable-wavelength three-dimensional reciprocal space mapping. This approach gave access to the tilt and strain distribution of Si micro-pillars during mechanical compression. The good agreement between results using rocking-curve-based and variable-wavelength quick scanning X-ray microscopy methods demonstrates that the latter preserves the strain and tilt-sensitive imaging capabilities of the first one without the need for sample motion. The technique can also be used to evaluate the composition and thickness fluctuations of thin structures.[22,23] Despite the chromaticity of Fresnel zone plates, the energy of the incident X-ray beam can be scanned by several tens of electron-volt without deteriorating the focal spot on the sample. To increase the scannable energy range, *i.e.* to achieve a larger reciprocal space range, during variable-wavelength quick scanning X-ray microscopy, the position of the Fresnel zone plate along the beam can be adjusted matching the energy-dependent focal distance. This experimental technique paves the way to novel combinations of *in situ* strain measurements of poly- and single-crystalline structures in heavy and complex environments. The detector can be positioned very close to the sample to catch most of the reflections of the grains. The energy-tuning approach circumvents both the comparatively large sphere of confusion of diffractometers compared with nanostructures and vibrations induced by motors. As the technique provides microscopy images with 3D reciprocal space information with a spatial resolution given by the nanobeam size, it will definitely benefit from the development of novel lenses. Recently, a record resolution leading to a X-ray spot size of 8.4 nanometers by 6.8 nanometers has been achieved.[36] The quick scanning technique will also benefit from the increase of the brilliance of X-rays from the multiple upgrade projects currently being carried out or planned at several third-generation synchrotron sources.

**Supporting Information**
Supporting Information is available from the Wiley Online Library or from the author.

**Acknowledgements**

The authors gratefully acknowledge the financial support from the French National Research Agency through the project ANR12-BS04-0003 BiDuL and from the German Science Foundation through the project 316662945. This work partially pertains to the French Government program "Investissements d'Avenir" (LABEX INTERACTIFS, reference ANR-11-LABX-0017-01) and has been partially supported by "Nouvelle Aquitaine" Region and by European Structural and Investment Funds (ERDF reference: P-2016-BAFE-94/95). We acknowledge the European Synchrotron Facility for beamtime allocation and we would like to thank for assistance in using beamline ID01. We thank Guillaume Amiard for the help during FIB sessions and Vincent Favre-Nicolin (ESRF) for fruitful discussions.

Received: ((will be filled in by the editorial staff))
Revised: ((will be filled in by the editorial staff))
Published online: ((will be filled in by the editorial staff))

**References**

[1] J. Li, Z. Shan, E. Ma, **2014**, *39*, 108.
[2] M. J. Hÿtch, J.-L. Putaux, J.-M. Pénisson, *Nature* **2003**, *423*, 270.
[3] M. J. Hÿtch, A. M. Minor, *MRS Bulletin* **2014**, *39*, 138.
[4] J. Li, Z. Shan, E. Ma, *MRS Bulletin* **2014**, *39*, 108.
[5] A. D. Krawitz, T. M. Holden, *MRS Bulletin* **1990**, *15*, 57.
[6] I. Robinson, R. Harder, *Nat Mater* **2009**, *8*, 291.
[7] M. Holt, R. Harder, R. Winarski, V. Rose, *Annual Review of Materials Research* **2013**, *43*, 183.
[8] G. A. Chahine, M.-I. Richard, R. A. Homs-Regojo, T. N. Tran-Caliste, D. Carbone, V. L. R. Jacques, R. Grifone, P. Boesecke, J. Katzer, I. Costina, H. Djazouli, T. Schroeder, T. U. Schülli, *J Appl Cryst, J Appl Crystallogr* **2014**, *47*, 762.
[9] P. Godard, G. Carbone, M. Allain, F. Mastropietro, G. Chen, L. Capello, A. Diaz, T. H. Metzger, J. Stangl, V. Chamard, *Nat. Commun.* **2011**, *2*, DOI 10.1038/ncomms1569.
[10] P. G. Evans, D. E. Savage, J. R. Prance, C. B. Simmons, M. G. Lagally, S. N. Coppersmith, M. A. Eriksson, T. U. Schülli, *Advanced Materials* **2012**, *24*, 5217.
[11] T. Etzelstorfer, M. Dierolf, G. Schiefler, V. Jacques, D. Carbone, D. Chrastina, G. Isella, R. Spolenak, J. Stangl, H. Sigg, A. Diaz, *J. Synchrotron Radiation* **2013**.
[12] Z. Ren, F. Mastropietro, A. Davydok, S. Langlais, M.-I. Richard, J.-J. Furter, O. Thomas, M. Dupraz, M. Verdier, G. Beutier, P. Boesecke, T. Cornelius, *Journal of Synchrotron Radiation* **2014**, *21*, 1128.
[13] M. Dupraz, G. Beutier, T. W. Cornelius, G. Parry, Z. Ren, S. Labat, M.-I. Richard, G. A. Chahine, O. Kovalenko, M. De Boissieu, E. Rabkin, M. Verdier, O. Thomas, *Nano Lett.* **2017**, *17*, 6696.
[14] T. W. Cornelius, D. Carbone, V. L. R. Jacques, T. U. Schülli, T. H. Metzger, *J Synchrotron Radiat* **2011**, *18*, 413.
[15] T. W. Cornelius, A. Davydok, V. L. R. Jacques, R. Grifone, T. Schulli, M.-I. Richard, G. Beutier, M. Verdier, T. H. Metzger, U. Pietsch, others, *Journal of synchrotron radiation* **2012**, *19*, 688.
[16] W. Cha, W. Liu, R. Harder, R. Xu, P. H. Fuoss, S. O. Hruszkewycz, *J Synchrotron Rad, J Synchrotron Radiat* **2016**, *23*, 1241.
[17] W. Cha, A. Ulvestad, M. Allain, V. Chamard, R. Harder, S. J. Leake, J. Maser, P. H. Fuoss, S. O. Hruszkewycz, *Phys. Rev. Lett.* **2016**, *117*, 225501.
[18] C. Leclere, T. W. Cornelius, Z. Ren, O. Robach, J.-S. Micha, A. Davydok, O. Ulrich, G. Richter, O. Thomas, *J Synchrotron Rad* **2016**, *23*, 1395.
[19] G. A. Chahine, M.-I. Richard, R. A. Homs-Regojo, T. N. Tran-Caliste, D. Carbone, V. L. R. Jacques, R. Grifone, P. Boesecke, J. Katzer, I. Costina, H. Djazouli, T. Schroeder, T. U. Schülli, *Journal of Applied Crystallography* **2014**, *47*, 762.
[20] O. Mandula, M. Elzo Aizarna, J. Eymery, M. Burghammer, V. Favre-Nicolin, *Journal of Applied Crystallography* **2016**, *49*, 1842.
[21] F. Mastropietro, A. Diaz, D. Carbone, J. Eymery, A. Sentenac, T. H. Metzger, V. Chamard, V. Favre-Nicolin, *Optics Express* **2011**, *19*, 19223.
[22] C. Kirchlechner, J. Keckes, J.-S. Micha, G. Dehm, *Adv. Eng. Mater.* **2011**, *13*, 837.
[23] C. Kirchlechner, W. Grosinger, M. W. Kapp, P. J. Imrich, J.-S. Micha, O. Ulrich, J. Keckes, G. Dehm, C. Motz, *Philosophical Magazine* **2012**, *92*, 3231.
[24] C. Ponchut, J. M. Rigal, J. Clément, E. Papillon, A. Homs, S. Petitdemange, *J. Inst.* **2011**, *6*, C01069.
[25] C. Neumann, S. Reichardt, P. Venezuela, M. Drögeler, L. Banszerus, M. Schmitz, K. Watanabe, T. Taniguchi, F. Mauri, B. Beschoten, S. V. Rotkin, C. Stampfer, *Nature Communications* **2015**, *6*, 8429.
[26] T. Lee, F. A. Mas'ud, M. J. Kim, H. Rho, *Scientific Reports* **2017**, *7*, 16681.
[27] B. Pan, K. Qian, H. Xie, A. Asundi, *Meas. Sci. Technol.* **2009**, *20*, 062001.
[28] A. J. Wilkinson, G. Meaden, D. J. Dingley, *Materials Science and Technology* **2006**, *22*, 1271.
[29] T. B. Britton, J. L. R. Hickey, *IOP Conference Series: Materials Science and Engineering* **2018**,

*304*, 012003.

[30]   T. C. Lee, I. M. Robertson, H. K. Birnbaum, *Philosophical Magazine A* **1990**, *62*, 131.

[31]   C. Gammer, J. Kacher, C. Czarnik, O. L. Warren, J. Ciston, A. M. Minor, *Applied Physics Letters* **2016**, *109*, 081906.

[32]   K. Müller, A. Rosenauer, M. Schowalter, J. Zweck, R. Fritz, K. Volz, *Microscopy and Microanalysis* **2012**, *18*, 995.

[33]   M. J. Hÿtch, A. M. Minor, *MRS Bulletin* **2014**, *39*, 138.

[34]   M. J. Hÿtch, T. Plamann, *Ultramicroscopy* **2001**, *87*, 199.

[35]   M.-I. Richard, M. Zoellner, G. Chahine, P. Zaumseil, G. Capellini, M. Häberlen, P. Storck, T. Schulli, T. Schroeder, *ACS Appl. Mater. Interfaces* **2015**, *7*, 26696.

[36]   S. Bajt, M. Prasciolu, H. Fleckenstein, M. Domaracký, H. N. Chapman, A. J. Morgan, O. Yefanov, M. Messerschmidt, Y. Du, K. T. Murray, V. Mariani, M. Kuhn, S. Aplin, K. Pande, P. Villanueva-Perez, K. Stachnik, J. P. Chen, A. Andrejczuk, A. Meents, A. Burkhardt, D. Pennicard, X. Huang, H. Yan, E. Nazaretski, Y. S. Chu, C. E. Hamm, *Light: Science & Applications* **2018**, *7*, 17162.

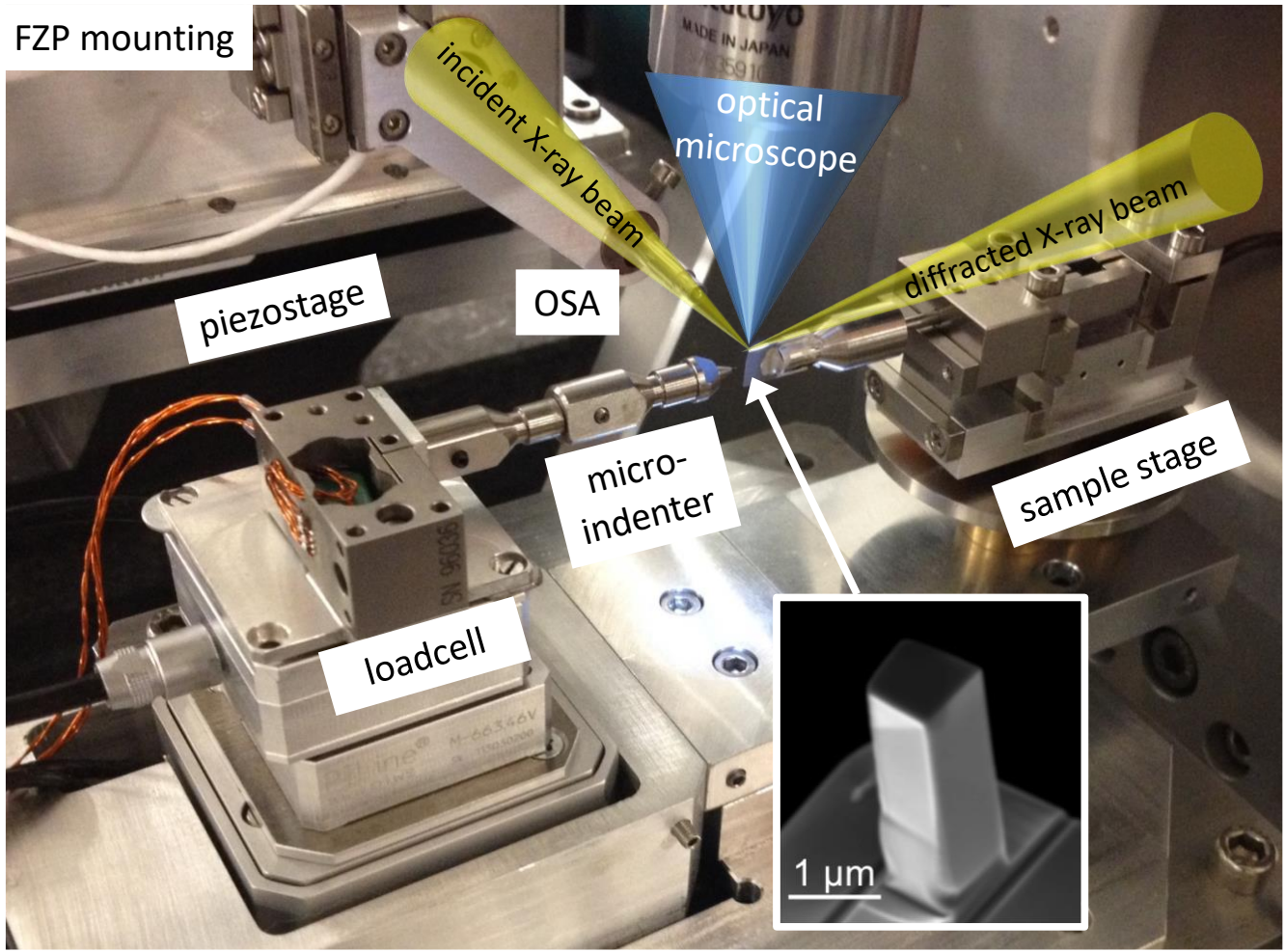

**Figure 1**. Experimental set-up showing (i) the load-cell with the sample and the diamond tip, (ii) the focusing optics (order sorting aperture (OSA) and box with a FZP inside) and (iii) translation motors (hexapod and piezo-motors). The inset displays one Si micro-pillar with height of 2.8 μm and lateral size of 0.9 μm.

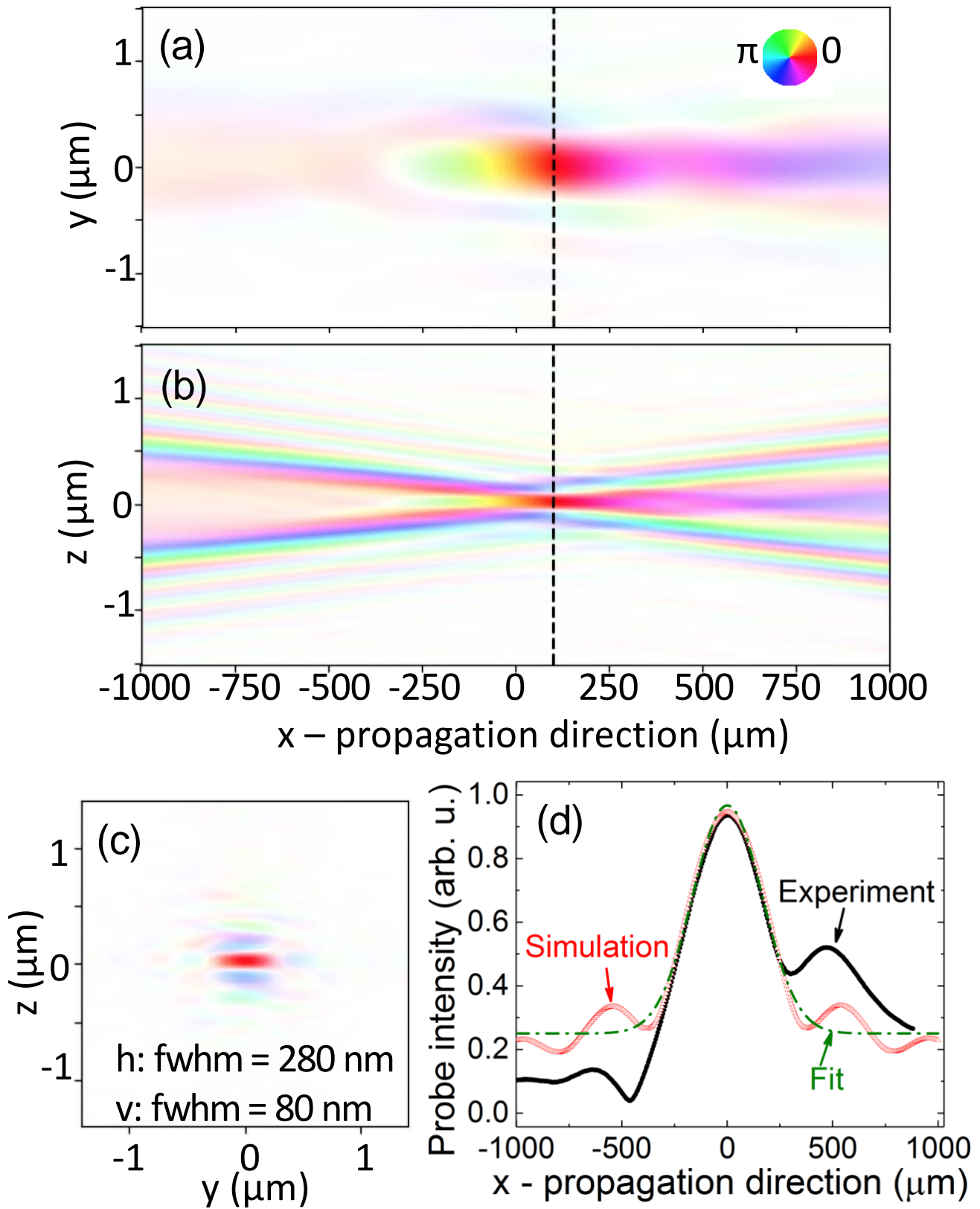

**Figure 2**. Experimental complex illumination in the direction of propagation, *x* and along *y* (a) and *z* (b). (c) Experimental complex illumination at the focal plane of the Fresnel Zone Plate. (d) Experimental

(black curve) and simulated (red curve) illumination profile in the direction of propagation. The green curve corresponds to a gaussian fit of the simulated curve.

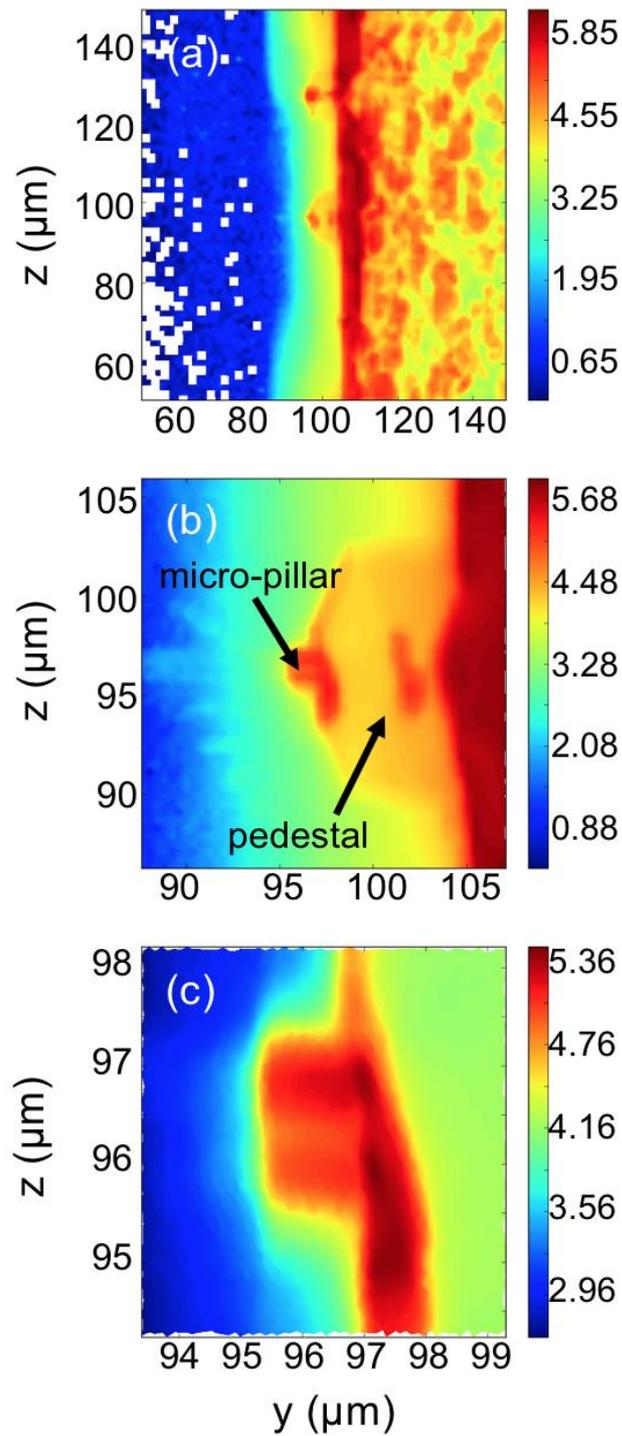

**Figure 3**. Two-dimensional direct space maps of the logarithmic integrated intensity measured at the **220** Si Bragg reflection for different surface areas (from large to small area (a-c)) centered on a Si micro-pillar.

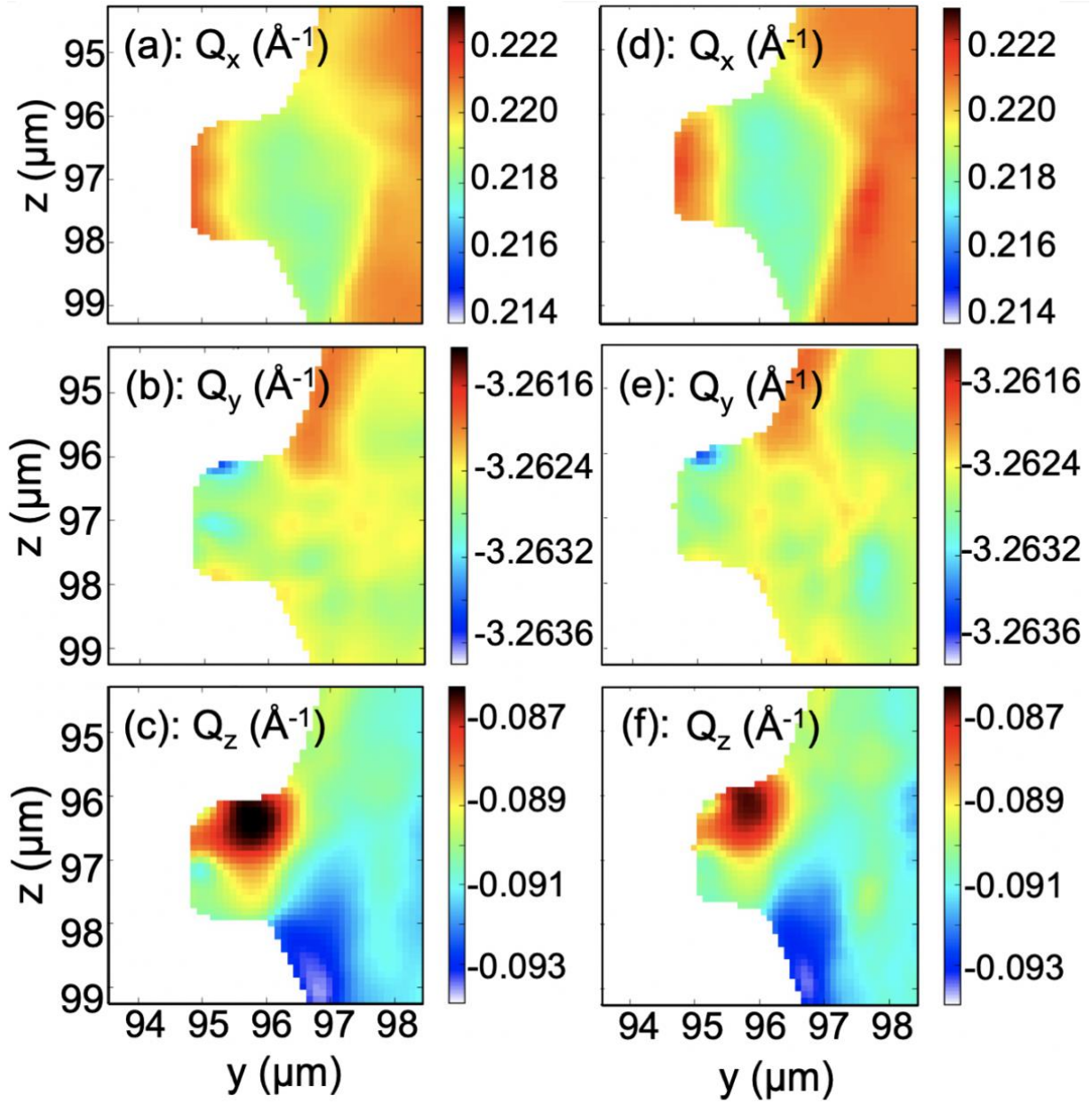

**Figure 4.** Reciprocal space coordinates of the scattering vector $Q_x$ (a, d), $Q_y$ (b, e) and $Q_z$ (c, f). Results obtained for the pristine pillar measured using the conventional (left: a,sb and c) and the variable-wavelength (right: d, e and f) quick scanning techniques.

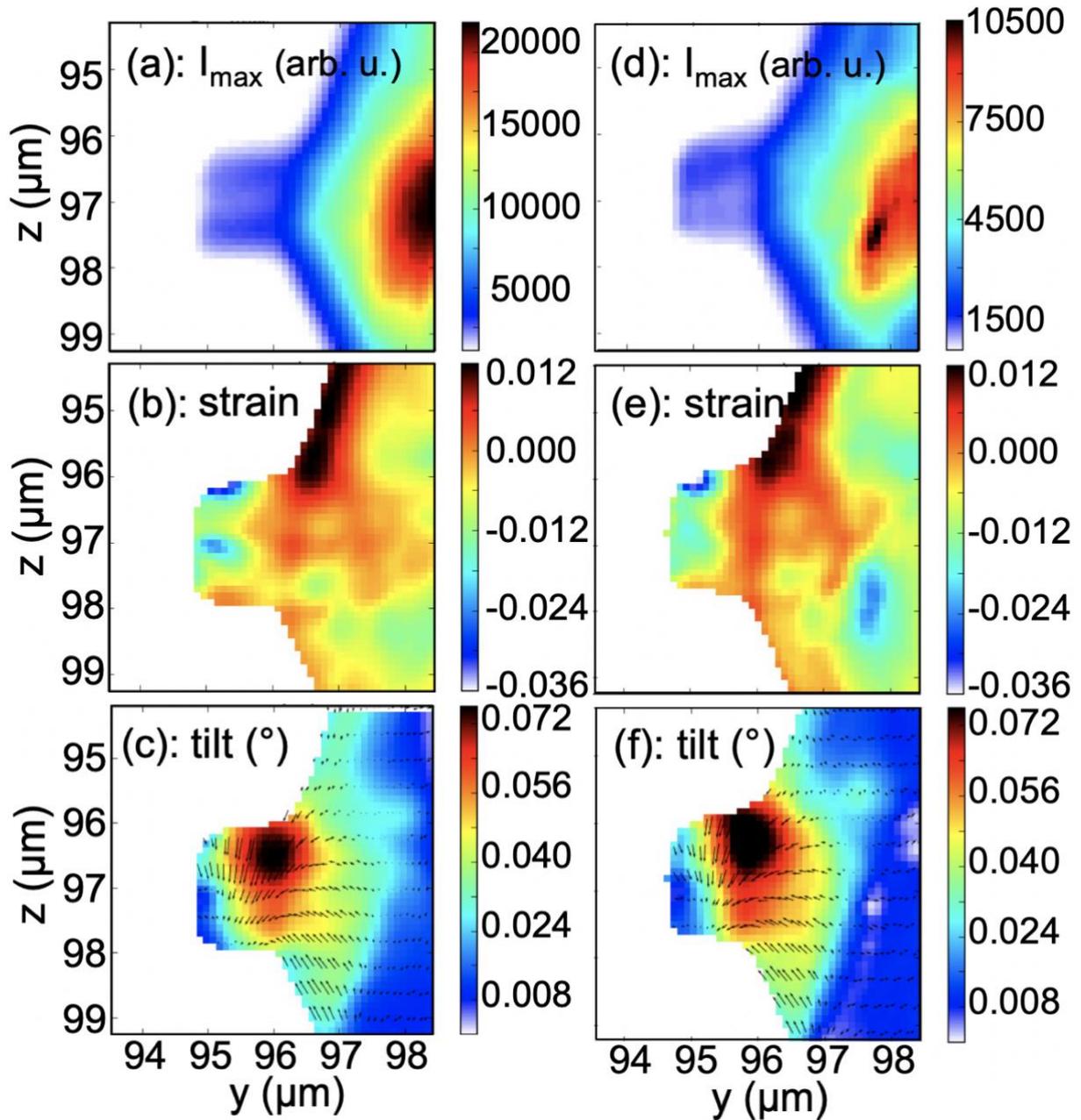

**Figure 5**. (a, d) Two-dimensional direct space maps of the maximum intensity measured at the **220** Si Bragg reflection. Maps of (b, e) the strain (in %) along the [220] direction and of (c, f) the tilt. Results obtained for the pristine pillar measured using the conventional (left: a, b and c) and the variable-wavelength (right: d, e and f) quick scanning techniques.

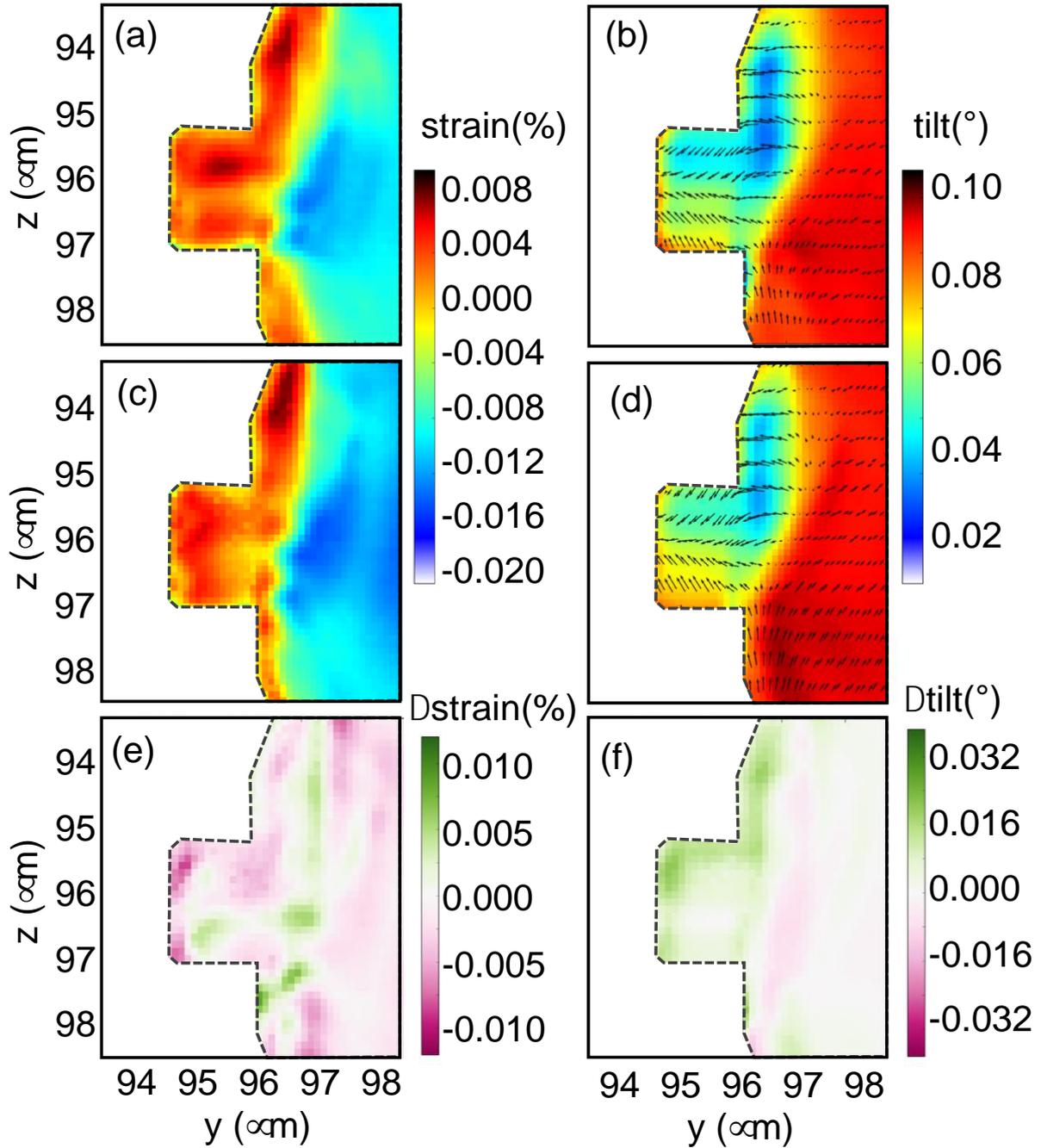

**Figure 6**. Maps of the (a-c) strain (in %) along the [220] direction and of the (b-d) tilt of the (220) lattice planes of a Si pillar and pedestal (shown in Figure 3 and different from Figures 4 and 5) before (a-b) and when the tip/indenter (displayed in grey) is in contact with the Si pillar (c-d). The differences of strain and tilt between the contact and the initial state are displayed in (e) and (f), respectively. The gray dashed-line indicates the contour of the Si pillar and the pedestal.

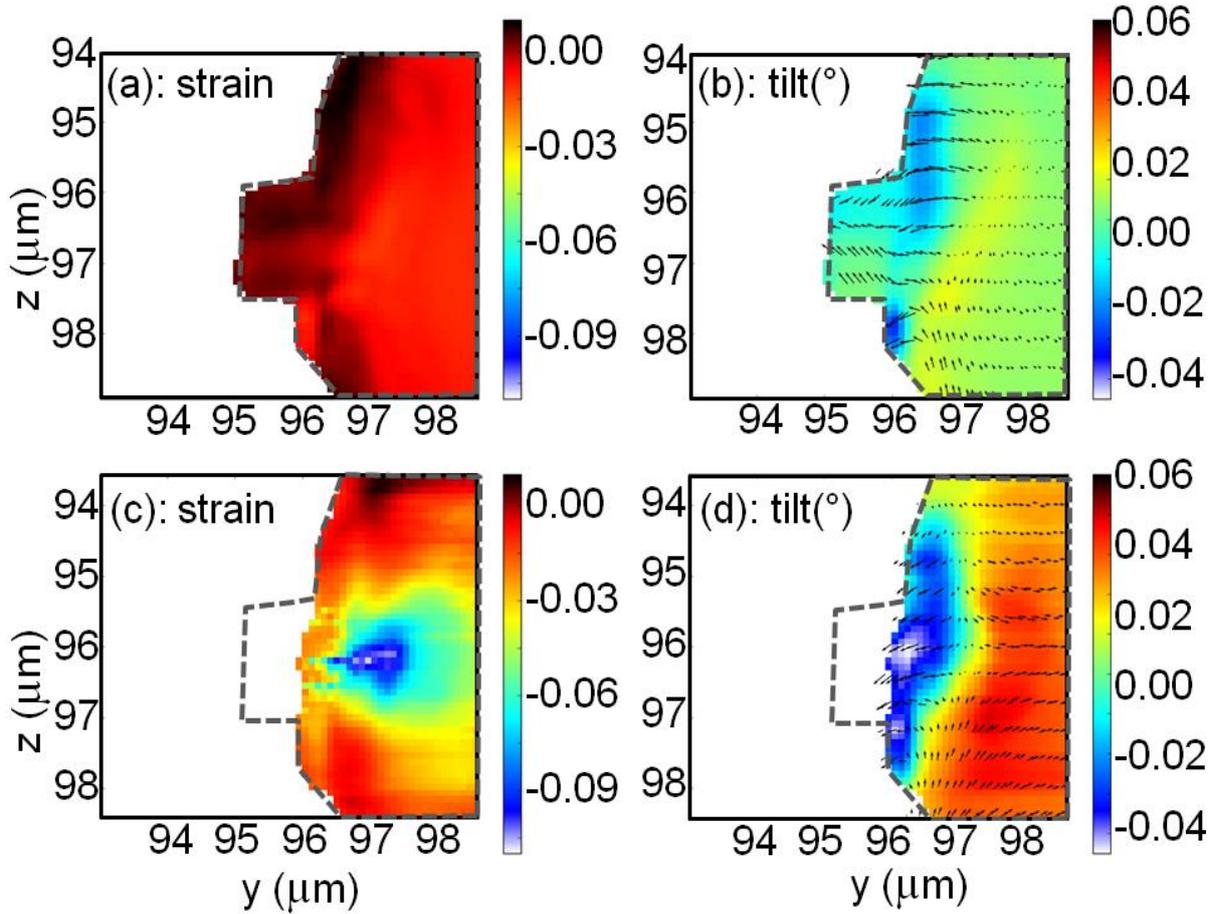

**Figure 7**. Maps of the (a-c) strain (in %) along the [220] direction and of the (b-d) tilt of the (220) lattice planes of a Si pillar and/or pedestal (shown in Figures 3 and 6) before (a-b) and during loading (c-d). The gray dashed-line indicates the contour of the Si pillar and the pedestal.

**Table of contents: Variable-wavelength quick scanning X-ray microscopy** is demonstrated to be a new *in situ* tool that provides access both to the rotation of the crystalline lattice and to the strain field inside micro- and nanostructures under mechanical load. It offers news opportunities to study the mechanical behavior at small scales *in situ* as well as in heavy and complex environments.

Keyword: structural microscopy, strain, lattice tilt, *in situ*, energy-scanning

Marie-Ingrid Richard[†,‡,*], Thomas W. Cornelius[†], Florian Lauraux[†], Jean-Baptiste Molin[§], Christoph Kirchlechner[§], Steven J. Leake[‡], Jérôme Carnis[†,‡], Tobias U. Schülli[‡], Ludovic Thilly[⊥], Olivier Thomas[†]

**Title: Variable-wavelength quick scanning X-ray microscopy for *in situ* strain and tilt mapping**

ToC figure:

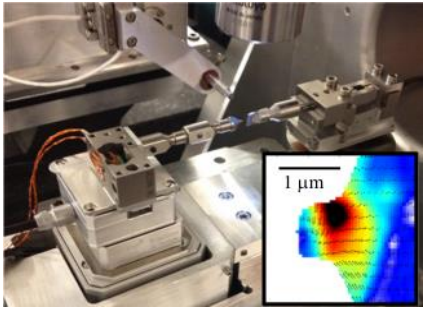